\documentclass[apj]{emulateapj}
\usepackage{natbib}
\usepackage{epstopdf}
\usepackage{CJK}
\usepackage{color}
\usepackage{pshan}
\usepackage{amsmath}
\newcommand{\beq}{\begin{equation}}
\newcommand{\eeq}{\end{equation}}
\newcommand{\bea}{\begin{eqnarray}}
\newcommand{\eea}{\end{eqnarray}}

\begin{document}
\begin{CJK}{UTF8}{mj}
\title{Alcock-Paczynski Test with the Evolution of Redshift-Space Galaxy Clustering Anisotropy}

\author{Hyunbae Park (박현배)\altaffilmark{1,2,3}}
\author{Changbom Park \altaffilmark{2}}
\author{Cristiano G. Sabiu\altaffilmark{4}}
\author{Xiao-dong Li\altaffilmark{5}}
\author{Sungwook E. Hong (홍성욱)\altaffilmark{6}}
\author{Juhan Kim (김주한)\altaffilmark{7}}
\author{Motonari Tonegawa\altaffilmark{2}}
\author{Yi Zheng\altaffilmark{2}}

\altaffiltext{1}{Korea Astronomy and Space Science Institute, 776 Daedeok-daero, Yuseong-gu, Daejeon, 34055, Korea}
\altaffiltext{2}{School of Physics, Korea Institute for Advanced Study, 85 Heogiro, Dongdaemun-gu, Seoul, 02455, Korea}
\altaffiltext{3}{Kavli IPMU (WPI), UTIAS, The University of Tokyo, Kashiwa, Chiba 277-8583, Japan}
\altaffiltext{4}{Department of Astronomy, Yonsei University, 50 Yonsei-ro Seodaemun-gu, Seoul, 03722, Korea}
\altaffiltext{5}{School of Physics and Astronomy, Sun Yat-Sen University, Zhuhai 519082, China}
\altaffiltext{6}{Natural Science Research Institute, University of Seoul,
163 Seoulsiripdaero, Dongdaemun-gu, Seoul, 02504, Korea}
\altaffiltext{7}{Center for Advanced Computation, Korea Institute for Advanced Study, 85 Heogiro, Dongdaemun-gu, Seoul, 02455, Korea}

\begin{abstract} 

We develop an improved Alcock-Paczynski (AP) test method that uses the redshift-space two-point correlation function (2pCF) of galaxies. Cosmological constraints can be obtained by examining the redshift dependence of the normalized 2pCF, which should not change apart from the expected small non-linear evolution. An incorrect choice of cosmology used to convert redshift to comoving distance will manifest itself as redshift-dependent 2pCF. Our method decomposes the redshift difference of the two-dimensional correlation function into the Legendre polynomials whose amplitudes are modeled by radial fitting functions. Our likelihood analysis with this 2-D fitting scheme tightens the constraints on $\Omega_m$ and ${w}$ by $\sim 40\%$ compared to the method of \citet{Li_2016,Li_2017,Li_2018} that uses one dimensional angular dependence only. 
We also find that the correction for the non-linear evolution in the 2pCF has a non-negligible cosmology dependence, which has been neglected in previous similar studies by Li {\it et al.}. With an accurate accounting for the non-linear systematics and use of full two-dimensional shape information of the 2pCF down to scales as small as $5~h^{-1}{\rm Mpc}$ it is expected that the AP test with redshift-space galaxy clustering anisotropy can be a powerful method to constraining the expansion history of the universe.

\end{abstract}

\section{Introduction}

\begin{figure*}
  \begin{center}
    \includegraphics[scale=0.65]{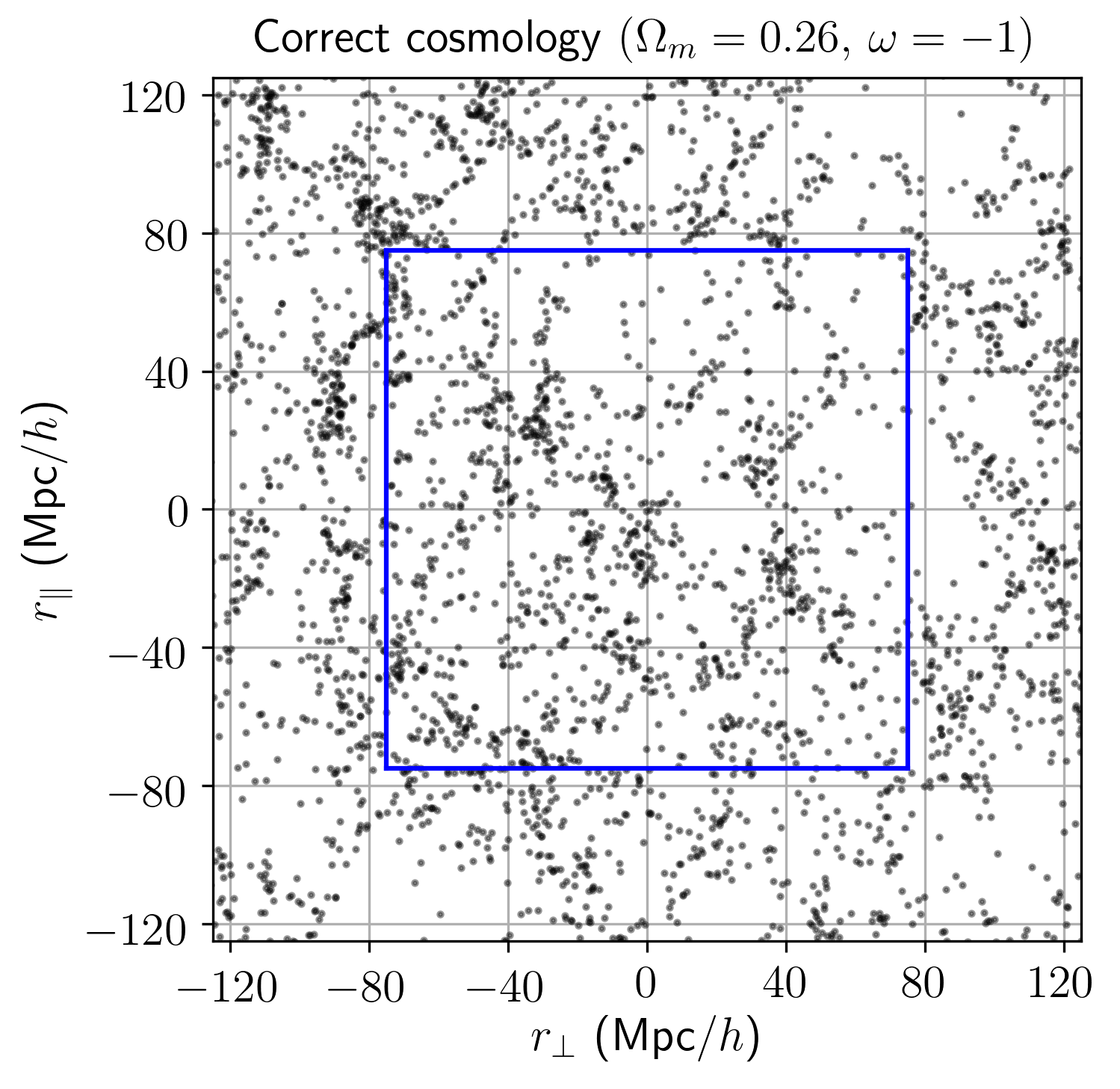}
    \includegraphics[scale=0.65]{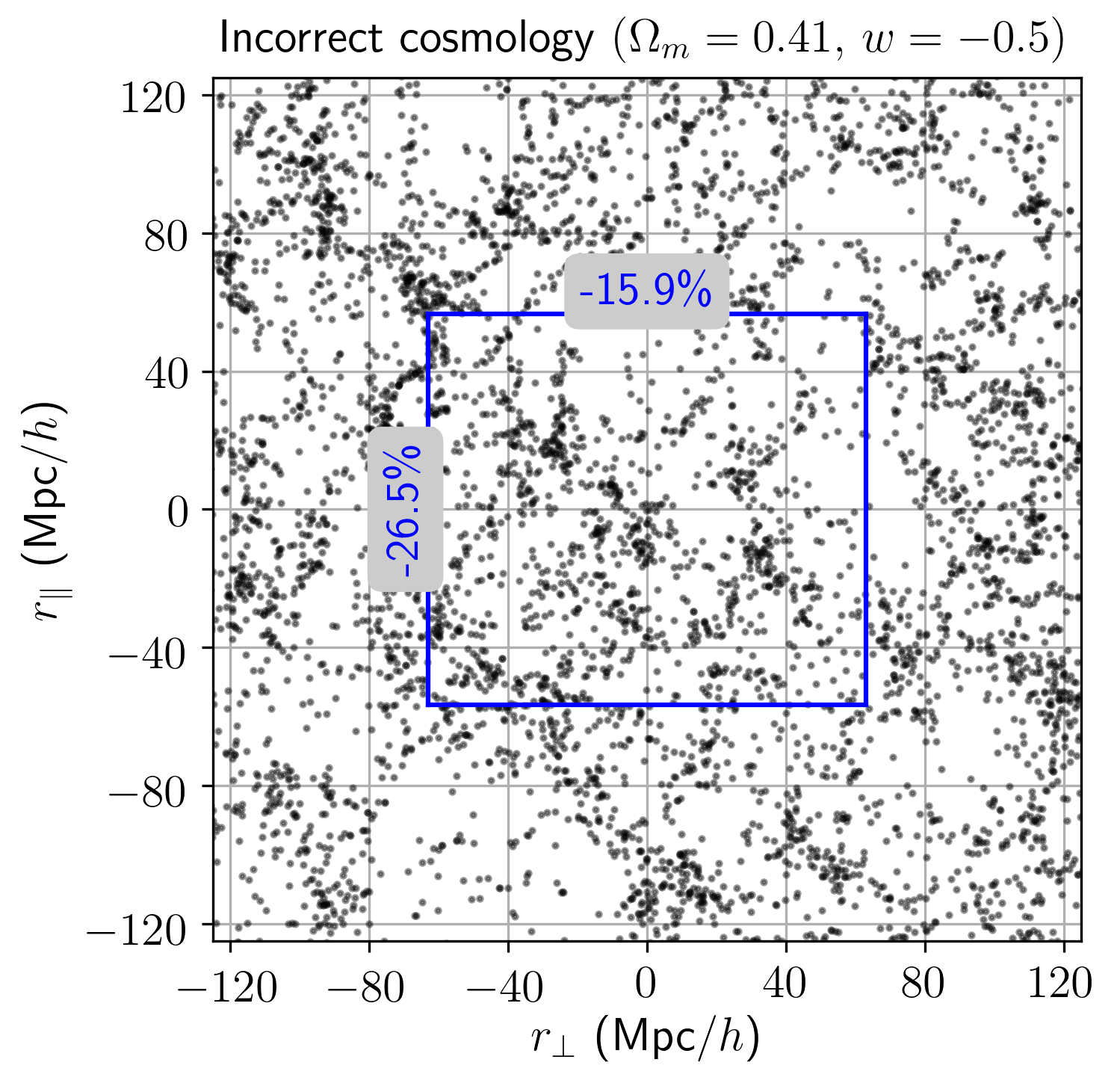}
  \caption{(left) Distribution of mock galaxies in the Horizon Run 4 simulation at $z=1$.
  (right) Same mock galaxies 
  with their distribution distorted by assuming $\Omega_m=0.46$ and $w=-0.5$ while the true cosmology has $\Omega_m=0.26$ and $w=-1$. }
  \label{fig:slice}
  \end{center}
\end{figure*}

The accelerated expansion of the Universe remains a deep mystery yet to be solved by contemporary cosmology. The most popular cosmological model, $\Lambda$CDM, exquisitely fits the peaks of the cosmic microwave background \citep{2013ApJS..208...19H,2018arXiv180706205P}, the distance-redshift relation of distant SNe1a \citep{1999ApJ...517..565P} and the distribution of galaxies \citet{Li_2016}. All the more reassuring is the small set of parameters required and the simplicity in the underlying assumptions of homogeneity, Gaussianity and near scale invariance of the initial perturbation spectrum. 

However, we are still left with the unsavoury prospect that if we are to believe $\Lambda$CDM then we are forced to include within our cosmic inventory a vacuum energy that is 120 orders of magnitude smaller than theoretical predictions \citep[e.g.,][]{weinberg1989} and a large component of matter that is not contained within the standard model of particle physics \citep[e.g.,][]{1987ARA&A..25..425T}. 

This has led many theorists to consider alternative models that include scalar field remnant from the Big Bang (see \citet{li2011} for review) and modifications to Einstein’s general relativity (see \citet{koyama2016} for review). Thus the endeavour of cosmology today does not lack a rich variety of theoretical models, but rather we lack precise observations of the expansion of the Universe, which would allow us discern among the proposed models.

Redshift survey is one of the most successful ways to obtain the data for uncovering the underlying cosmology of the universe. Many statistical tools have been developed to extract information on the initial conditions of the cosmic structures from observed spatial distribution of galaxies. The size and depth of surveys have been improved thanks to technological advances. Upcoming and ongoing surveys such as DESI \citep[Dark Energy Spectroscopic Instrument;][]{flaugher2014}, eBOSS \citep{2016ApJ...833..225Z},
and PFS \citep[Subaru Prime Focus Spectroscopy;][]{2014PASJ...66R...1T} 
will measure locations of millions of galaxies up to redshift of two or more together with many of those at redshifts below 1.

Clustering of galaxies on very large scales is in the linear regime still keeping information of the early universe, and therefore can be directly compared with the predictions of theoretical models. If there are statistical measures of galaxy clustering that suffer from little non-linear evolution effects either because the scale under study is safely in the linear regime or because they are intrinsically insensitive to nonlinear effects, they will be very useful in uncovering the physics of the early universe. In addition, they can be used for reconstructing the expansion history of the universe. This is because the conversion of observed redshift to comoving distance can create artificial systematic distortion of galaxy clustering when the cosmology adopted for the conversion is incorrect. \citet{2010ApJ...715L.185P} has adopted this idea and proposed to use the shapes of the 2pCF, power spectrum, or the genus topology of large-scales structures in the universe in particular as the cosmic invariants for constraining the cosmological parameters governing the expansion history of the universe \citep[see also,][]{2017ApJ...836...45A, 2018ApJ...853...17A}.


The Alcock-Paczinsky (AP) test is one of the statistical means to extract cosmological parameters from galaxy redshift survey data \citep[][see Figure~\ref{fig:slice}]{alcock1979}. It is a test of the geometry of cosmic objects or galaxy distribution, which should appear isotropic at all redshifts if the objects and galaxy clustering are spherical or isotropic and redshift of galaxy is converted to distance from correct cosmology. 
For the test, the Baryonic Acoustic Oscillation (BAO) feature in the galaxy 2pCF is often used due to its distinct excess at $r\sim 100~h^{-1}~\rm Mpc$. The AP test with the BAO has been proven to put a moderately powerful constraint of the parameters like the matter density parameter $\Omega_m$, dark energy equation of state $w$, and so on. A weakness of the BAO method is the fact that it uses the clustering information on very large scales where the statistics of a given sample is relatively weak \citep{Li_2016}. 

An extension of the AP test came from the observation that, even though the observed galaxy clustering in the redshift space is quite anisotropic due to the redshift-space distortion effects, its redshift evolution is almost conserved with only small non-linear effects \citep{Li_2015}. Redshift evolution of the shape of the 2pCF turned out to be a powerful tool for uncovering the expansion history of the universe as the signal of the 2pCF is much stronger near $10~h^{-1}$Mpc scale than at the BAO scale of $\sim 100~h^{-1}$Mpc due to much larger number of galaxy pairs. It has been shown that the shape of the redshift-space 2pCF down to $6~h^{-1}$Mpc scale does not evolve as much as incorrect cosmology assumption would distort it, making it possible to separate the systematic effects due to adopting incorrect cosmology \citep{Li_2015,Li_2016,Li_2018,2017ApJ...836...45A, 2018ApJ...853...17A, 2018ApJ...863..200A}.

There is another stream of efforts by \citet{2019A&A...621A..69R} that tackles this problem using Bayesian inference framework. Their work also aims to improve the cosmological constraint by utilizing 2-point statistics of galaxies at the entire range of scales.

This work is a continuation of our effort for improving the AP test with galaxy clustering anisotropy. Previous works by \citet{Li_2016, Li_2017, Li_2018} have left rooms for improvement of the method, some of which we attempt to accommodate in this work. The main contribution of this paper is as follows.

(1) \citet{Li_2016, Li_2017, Li_2018} used the angular dependence of the radially integrated 2pCF, which potentially dilutes the constraints from the radial shape of the 2pCF. We shall attempt to use both angular and radial shapes of the 2pCF to put the constraints and see how much the constraints improve. 

(2) A potential caveat in the previous work is that the systematic correction due to intrinsic evolution of the redshift-space 2pCF was assumed to be cosmology-independent without proof. Now that we have data set for multiple cosmologies (see Sec.~\ref{sec:multiverse} for detail), we can verify whether or not the assumption is correct.

The rest of the paper is as follows. In Section~\ref{sec:data}, we list and explain the $N$-body simulations and mock galaxy catalogs used for our analysis. We propose our new AP test method in detail in Section~\ref{sec:method}. We present the results in Section~\ref{sec:results}. Finally, we summarize and conclude in Section~\ref{sec:summary}.

\begin{figure*}
  \begin{center}
    \includegraphics[scale=0.6]{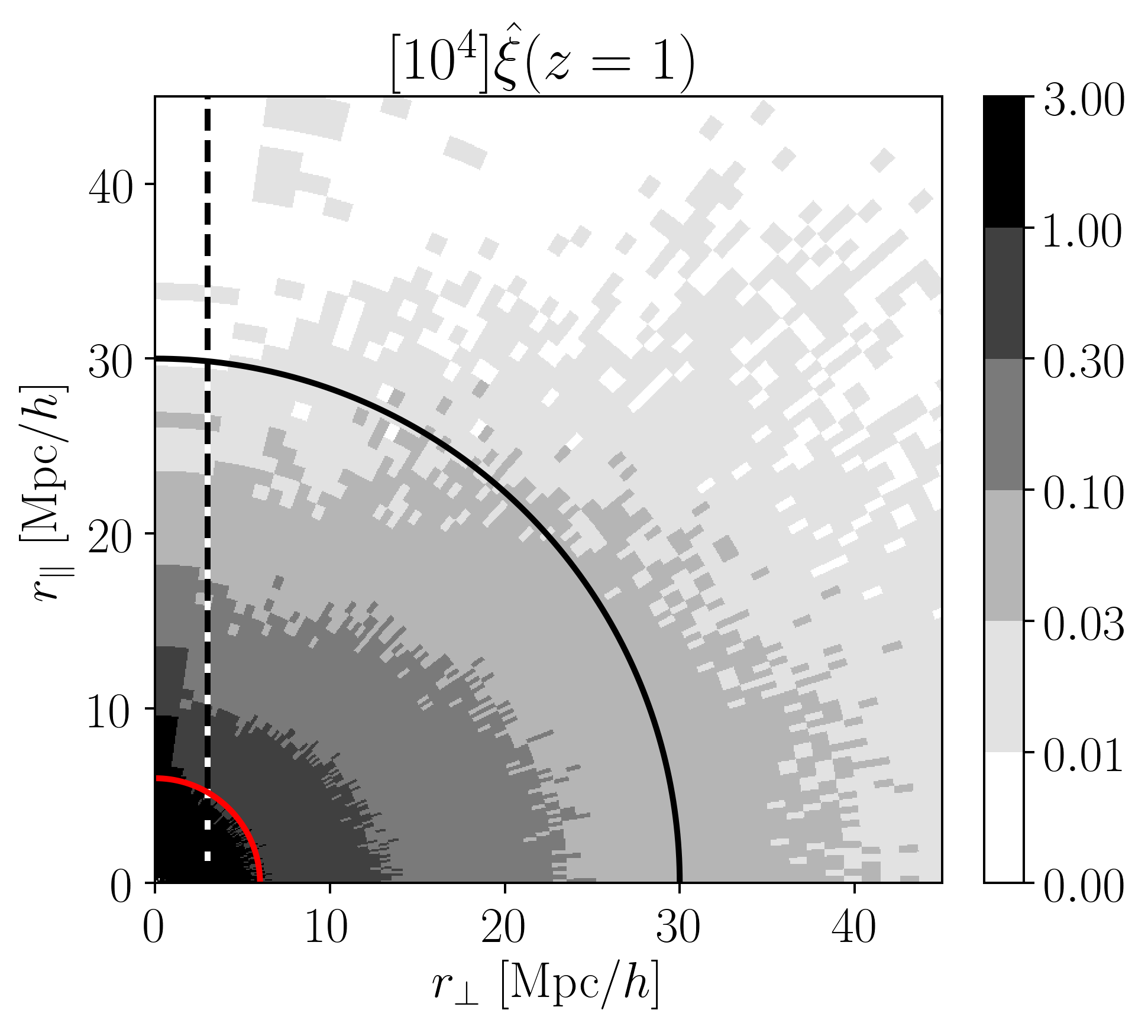}
    \includegraphics[scale=0.6]{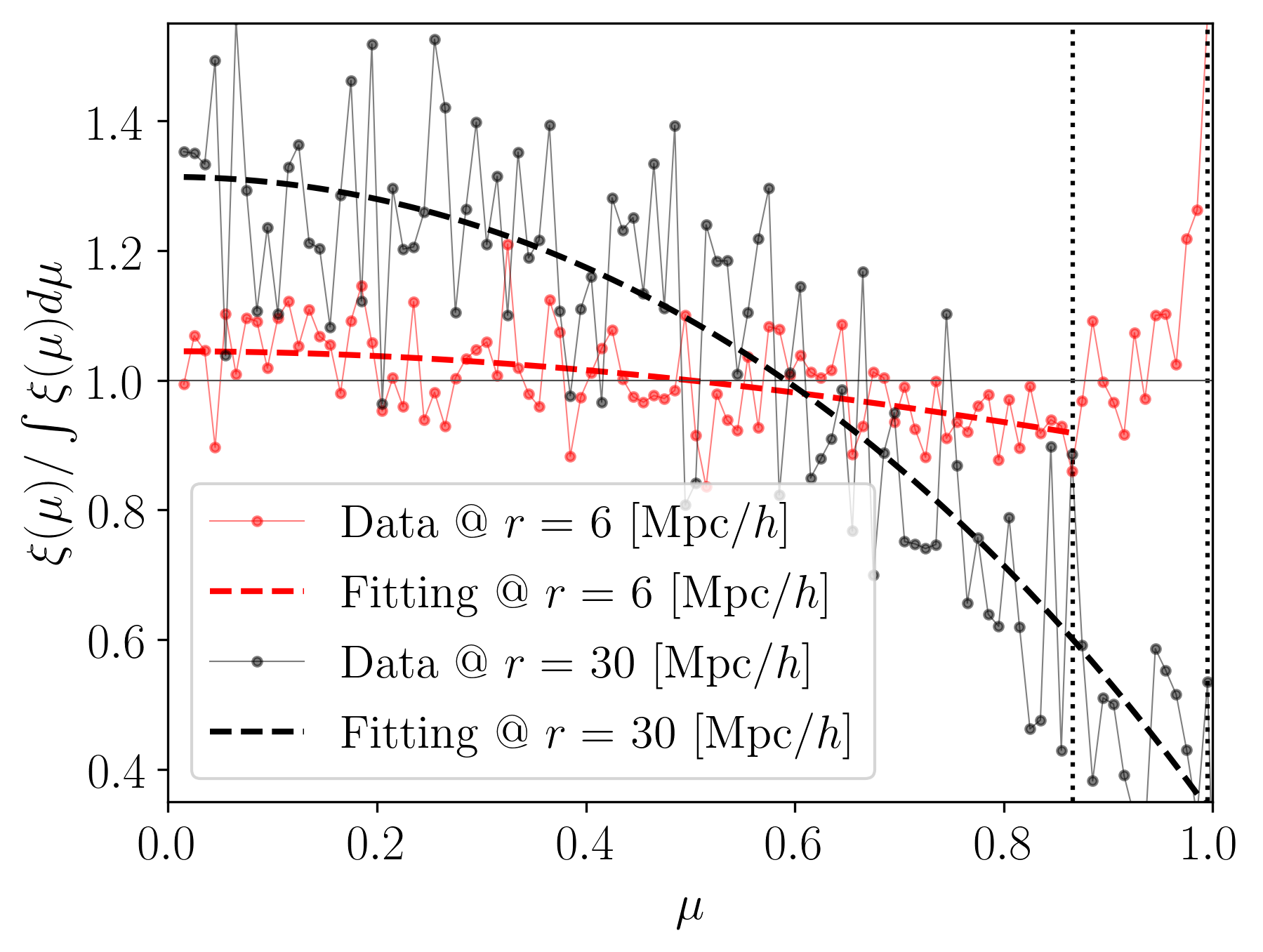}
  \caption{Left: Normalized correlation function $\hat\xi$ of mock galaxies at $z=1$. In the left of the vertical dashed line at $r_\perp = 3~h^{-1}{\rm Mpc}$ is the part that we exclude from our analysis. The arcs denote circles of $r=6$ (red) and $30~h^{-1}{\rm Mpc}$ (black), on which we show the angular shape of $\hat\xi$ in the right panel. Right: $\xi(r,\mu)$ is shown as a function of the cosine of the angle between the line-of-sight and the separation direction, $\mu$, for two pair-separations, $r=6$ (red) and $30~h^{-1}{\rm Mpc}$ (black). The dots and thin solid lines are the data points and the thick dashed lines are our fit for the data.}
  \label{fig:xi}
  \end{center}
\end{figure*}

\section{Data} \label{sec:data}
\subsection{Horizon Run 4} \label{sec:HR4}

The Horizon Run 4 (HR4) simulation \citep{2015JKAS...48..213K} is a massive cosmological simulation that evolved $N_p=6300^3$ particles in a cubic box of a side length of $L_{\rm box}=3150 ~h^{-1}{\rm Mpc}$. 
It uses a flat $\Lambda$CDM cosmological model in concordance with \emph{Wilkinson Microwave Anisotropy Probe} (WMAP) 5-year observation \citep{dunkley2009}, where the matter density fraction, dark energy density fraction, and dark energy equation of state at $z = 0$ is $(\Omega_m, \Omega_{de}, {w}) = (0.26,0.74,-1)$.
The volume of the HR4 is big enough to simulate the formation of large-scale structures, and at the same time its force and mass resolutions are high enough to simulate the formation of individual galaxies down to a relatively small mass scale. Thanks to these unique features, the HR4 has been extensively used for cosmological model tests and study of galaxy formation under the influence of large-scale structures in the universe.  \citep{2015JKAS...48..213K,2018MNRAS.477.2772U, 2018MNRAS.473.5098U,Li_2016,Li_2017, 2016ApJ...818..173H,2017ApJ...836...45A,2018ApJ...853...17A,2018A&A...620A.149E}. 

Rich information on structure formation is contained in the merger trees of dark matter (DM) halos forming in the big simulation box of HR4, constructed at 75 timesteps between $z = 12$ and 0 with the time interval of $\sim 0.1~{\rm Gyr}$.
In each snapshot, DM halos are found with the Friend-of-Friends  (FoF) algorithm with the linking length of $\ell_{\rm FoF} = 0.1~h^{-1} {\rm Mpc}$. 
The minimum number of DM particles constructing DM halos is set to 30, which corresponds to the minimum DM halo mass of $M_{\rm halo}^{\rm min} = 2.7 \times 10^{11}~h^{-1} {\rm M}_\sun$.
The mock galaxy catalogs of HR4 was modelled by applying the most bound halo particle (MBP)-galaxy abundance matching to its DM halo merger tree \citep{2016ApJ...823..103H}.
For each DM halo at each snapshot, we found the most gravitationally bound member particle (MBP).
The particle is marked as the center of a `galaxy' if the given halo is isolated or if it is the most massive member halo (namely the central halo) in the merger events.
On the other hand, for less massive member halos (satellites), we trace their `galaxies' from the time when they were isolated ones just before merger until they are completely disrupted.
The time between the infall and the complete disruption of satellite galaxies is estimated by adopting a modified merger timescale model of \citet{jiang2008}:
\begin{equation}
\frac{t_{\rm merge}}{t_{\rm dyn}} = \frac{(0.94 \epsilon^{0.60} + 0.60)/0.86}{\ln [1+(M_{\rm host}/M_{\rm sat})]}
\left( \frac{M_{\rm host}}{M_{\rm sat}} \right)^{\alpha} \, ,
\end{equation}
where $\epsilon, M_{\rm host}, M_{\rm sat}, t_{\rm dyn}$ are the circularity of the satellite's orbit, mass of central and satellite halos, and the orbital period of virialized objects, respectively. We set $\alpha=1.5$, which makes the 2pCF of our mock galaxies match that of the SDSS Main galaxies down to scales below $1~h^{-1} {\rm Mpc}$  \cite[]{zehavi2011}.


For our analysis, we divide the HR4 simulation box into 6 pieces in each dimension, thereby creating 216 sub-cube mock samples that are $525~h^{-1}{\rm Mpc}$ long on a side. This choice is made to have an enough number of samples for likelihood analysis. Some galaxy surveys like the SDSS cover a larger volume at the redshift of our interest ($z\sim1$). Thus, we plan to analyze larger sample volumes with larger simulations in future studies. 

We adopt $10^{-3}$ galaxy per $(h^{-1}{\rm Mpc})^3$ for the galaxy number density in the mock sample, which corresponds to 0.145 million in each sub-cube mock. This number density roughly correspond to the $r$-band magnitude $\mathcal{M}_r - 5 \log h < -20.3$ at $z = 0$ \citep{choi2010} and it is also similar to the expected number density galaxies to be observed by the PFS survey. We will also show some results with 10 times more galaxies for comparison. We note that these mass-cuts are rather arbitrary. The actual value to be used in the analysis of a given observational data should be determined by the survey data.

\begin{figure*}
  \begin{center}
    \includegraphics[scale=0.6]{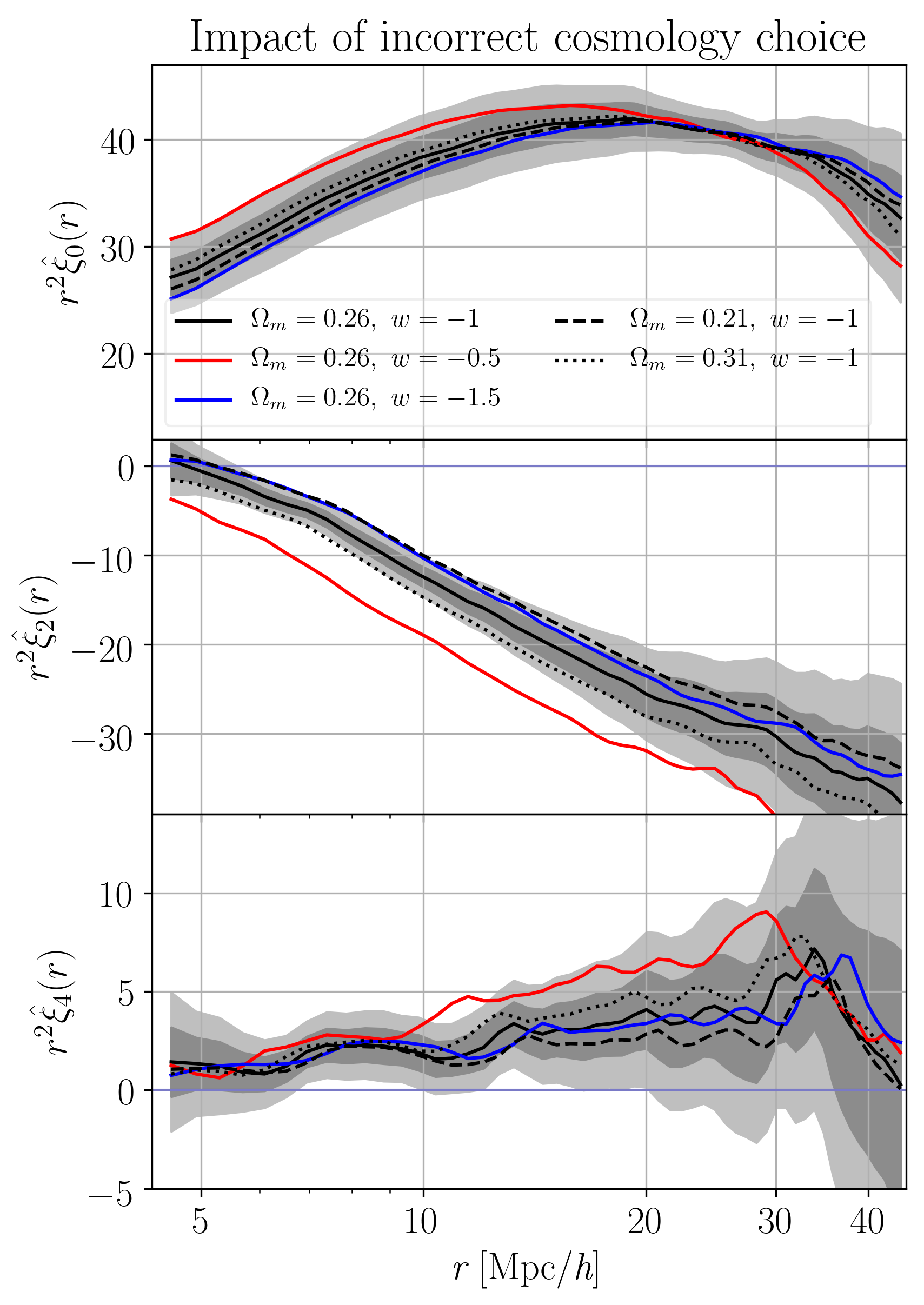}
    \includegraphics[scale=0.6]{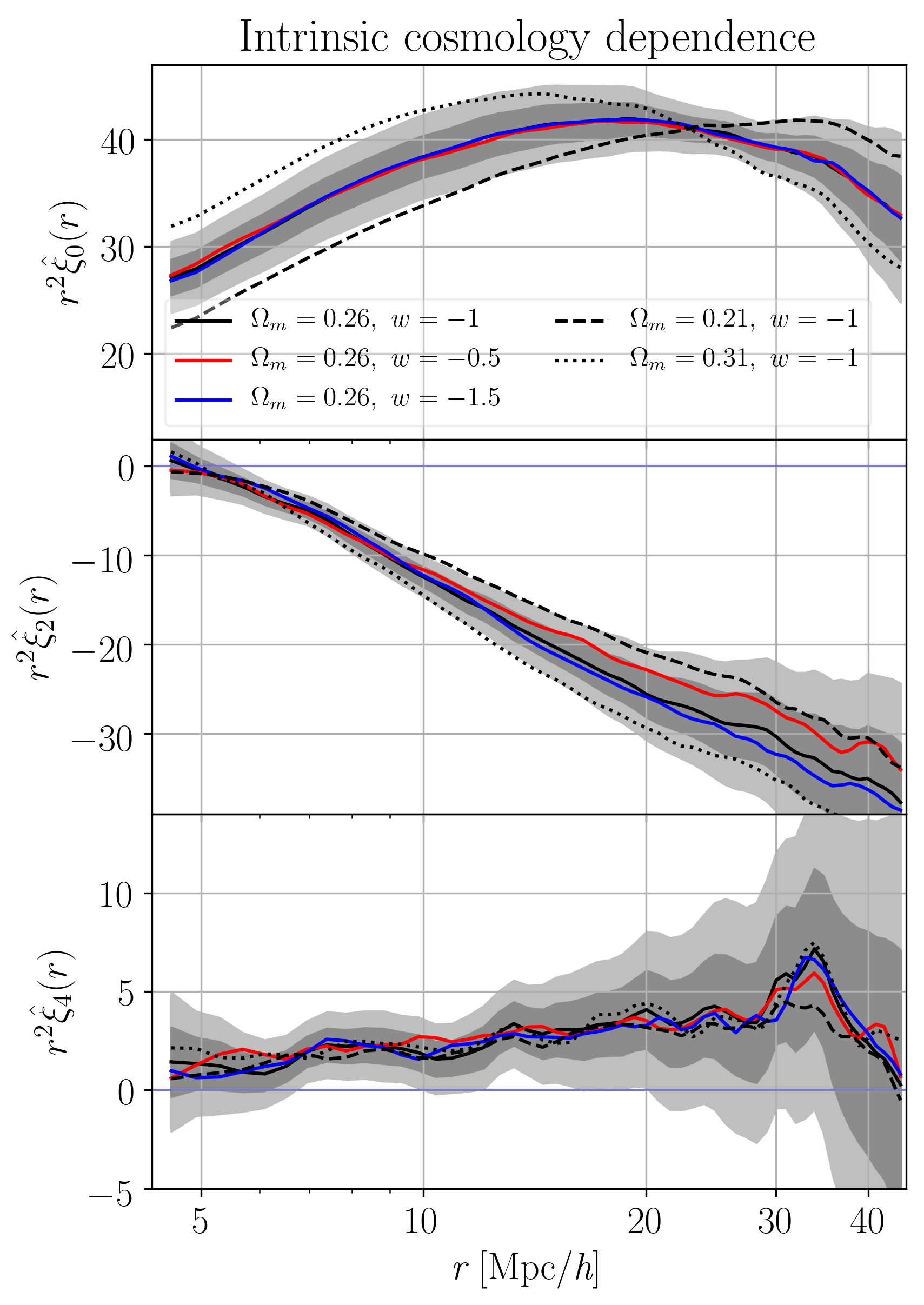}
  \caption{Left: $r^2 \hat\xi_0(r,z=1)$ (upper), $r^2\hat\xi_2(r,z=1)$ (middle), and $r^2 \hat\xi_4(r,z=1)$ (lower) for the fiducial model of HR4 is plotted at the black solid lines. Then the uncertainty range of those values for $(525$ $h^{-1}$ Mpc$)^3$ volume is shown in dark grey (1-$\sigma$ range) and grey (2-$\sigma$ range) shade based on the results from 216 sub-cube mock samples of HR4 with $10^{-3}$ galaxies per (Mpc/$h$)$^3$. Each sub-cube contains 0.145 million galaxies. The red, blue, black dashed, and black dotted lines are results of distorting the spatial galaxy distribution in the fiducial case by incorrectly assuming ($\Omega_m, {w}$) = ($0.26, -0.5$), ($0.26, -1.5$), ($0.21, -1$), and ($0.31, -1$), respectively. Right: The lines are the results of each of the Multiverse simulation set. The lines now describe the intrinsic shape of moments in each cosmology.}
  \label{fig:xi024}
  \end{center}
\end{figure*}

\subsection{Multiverse Simulations} \label{sec:multiverse}
\begin{table}[t]
\caption{Multiverse Simulation parameter}
\begin{center} 
\begin{tabular}{@{}llllllllllllllllllll||}  
\hline
Label & $w$  &  $\Omega_m$ &  $\Omega_{de}$  \\
\hline
Low-$w$      & $-1.5$ & 0.26 & 0.74 \\
Low-$\Omega_m$     & $-1$   & 0.21 & 0.79 \\
Fiducial       & $-1$   & 0.26 & 0.74 \\
High-$\Omega_m$     & $-1$   & 0.31 & 0.69 \\
High-$w$       & $-0.5$ & 0.26 & 0.74 \\
\hline
\end{tabular}  
\end{center}
\label{table: fitting}
\end{table}

The Multiverse simulations are a set of cosmological $N$-body simulations designed to see the effects of cosmological parameters on the clustering and evolution of cosmic structures.  We changed the cosmological parameters around those of the standard concordance model with $\Omega_m=0.26$, $\Omega_{de} = 0.74$, and $w=-1$. We used exactly the same set of random numbers to generate the initial density fluctuations of all the simulations, which allow us to make the proper comparison between the models with the effects of the cosmic variance compensated.

Five multiverse simulations we use in this paper are listed in Table 1. Two models have the matter density parameter shifted by 0.05 from the fiducial model while the dark energy equation of state is fixed to $w=-1$. The other two Quintessence models \citep{2011JCAP...03..047S} have $w$ shifted by 0.5 from the fiducial value of $-1$ while $\Omega_m$ is fixed to 0.26. These parameters are chosen so that they are reasonably large enough to cover the area in the $\Omega_m$-$w$ space constrained by many existing studies at the time WMAP 5-year results have been announced \citep{2003ApJS..148..175S}.


The power spectrum of each model is normalized in such a way that
the RMS of the matter fluctuation linearly evolved to $z=0$ has $\sigma_8=0.794$ when smoothed with a spherical top hat with $R=8~h^{-1} {\rm Mpc}$. 

The number of particles evolved is $N_p=2048^3$ and
the comoving size of the simulation box is $1024~h^{-1}{\rm Mpc}$. The starting redshift is $z_{\rm init}=99$ and the number of global time steps is 1980 with equal step size in the expansion parameter, $a$. We have used the CAMB package to calculate the power spectrum at $z_{\rm init}$.
We have extended the original GOTPM code \citep{2004NewA....9..111D} to gravitationally evolve particles according to the modified Poisson equation of
\begin{equation}
    \nabla^2\phi = 4\pi Ga^2 \bar{\rho}_m \delta_m\left(1 +{D_{de}\over D_m}{\Omega_{de}(a)\over\Omega_m(a)}\right),
\end{equation}
where $D_{de}$ and $D_m$ are the linear growth factors of the dark energy and matter, respectively (see \citealt{2011JCAP...03..047S} for details).

\section{Methodology} \label{sec:method}

\begin{figure*}
  \begin{center}
    \includegraphics[scale=0.6]{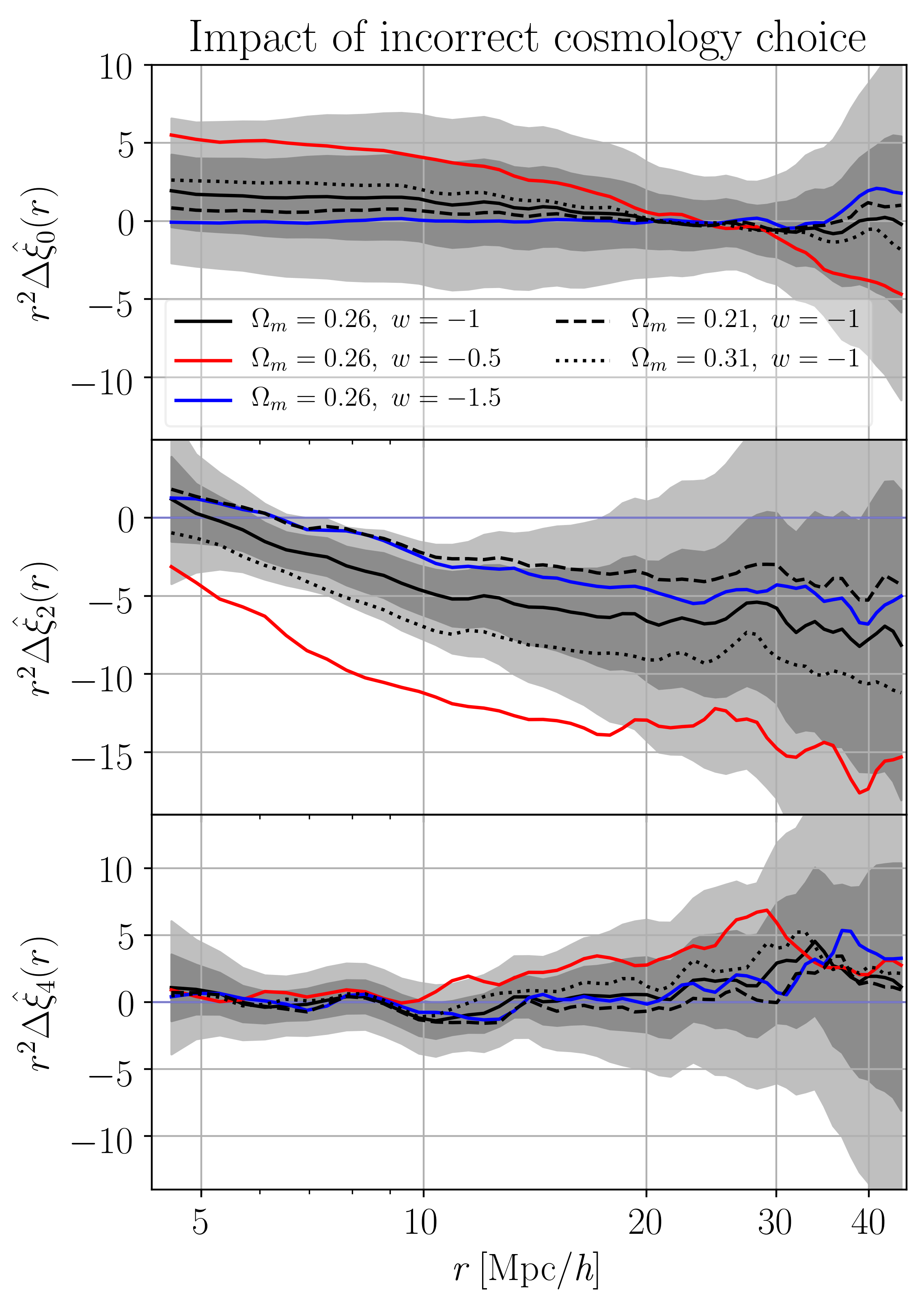}
    \includegraphics[scale=0.6]{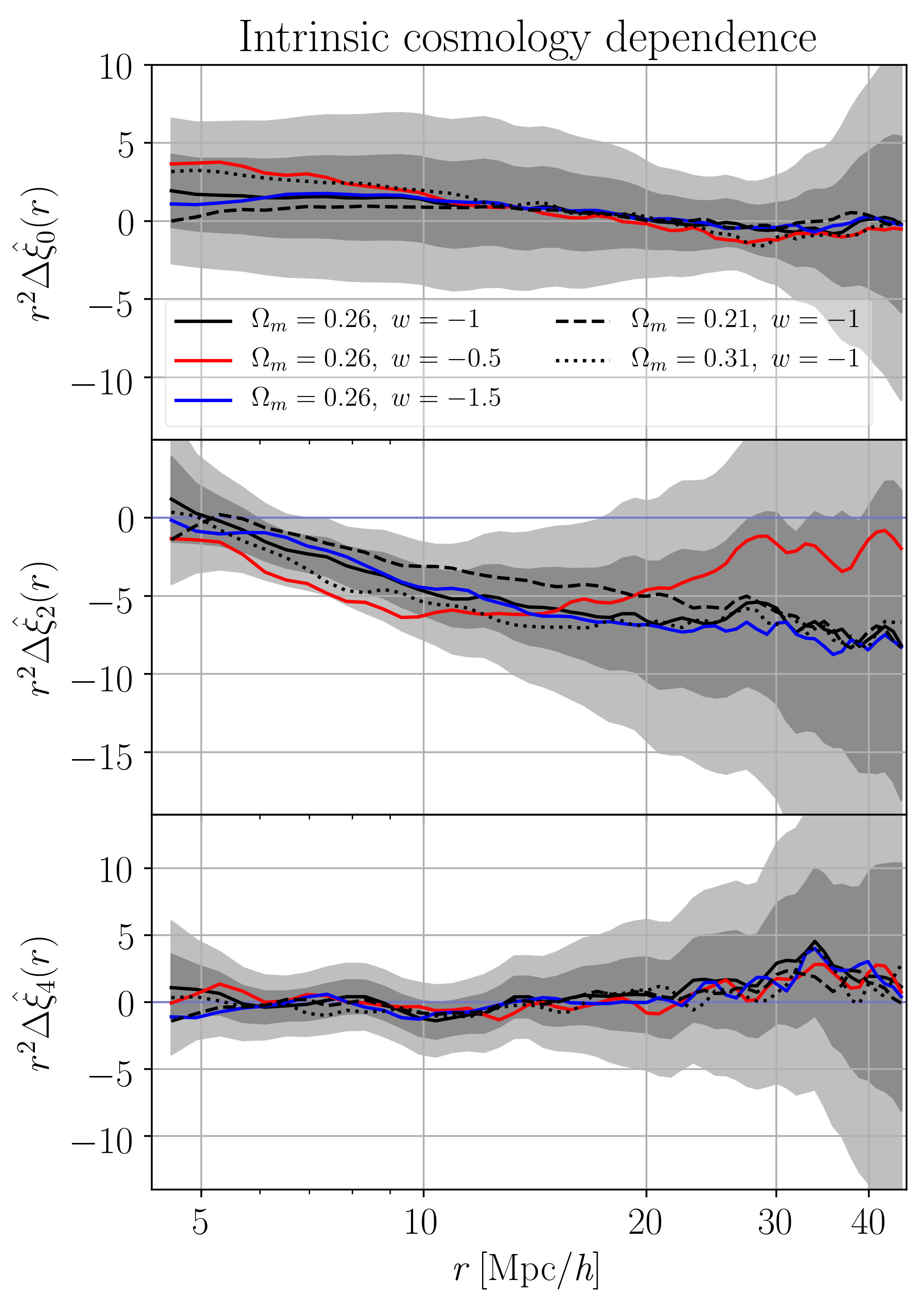}
  \caption{Similar to Figure~\ref{fig:xi024} except that we plot the redshift evolution between $z=0$ and 1,  $r^2\Delta\hat\xi_0(r,z_i=0,z_j=1)$, $r^2\Delta\hat\xi_2(r,z_i=0,z_j=1)$, and $r^2\Delta\hat\xi_4(r,z_i=0,z_j=1)$ in the upper, middle, and lower panels, respectively. The black solid lines show the non-linear evolution of each multipole between $z=0$ and 1 in the case of the fiducial cosmology, ($\Omega_m,{w}$)=(0.26,-1). On the right panels it is shown that the amount of redshift evolution depends weakly on cosmology.}
  \label{fig:delxi024}
  \end{center}
\end{figure*}

\subsection{Shape of 2-point Correlation Function}

In our AP test we use the two-dimensional shape of the galaxy 2pCF in the plane of the line-of-sight  and across the line-of-sight directions. However, we exclude the region $r_\perp<r_{\perp,\rm cut} \equiv 3~h^{-1}{\rm Mpc}$ of the plane from our analysis to minimize the impact of the highly nonlinear Finger-of-God effects. This leaves us with only a mildly non-linear contribution to $\xi$ in our analysis.

Since we use only the shape of $\xi$ and not its amplitude in the AP test, we shall normalize $\xi$ by its volume integral up to the radial separation of $r_{\rm max} = 45~h^{-1}~\rm{Mpc}$. Namely, the normalization factor is
\bea \label{eq:normalize1}
\xi_* (z) \equiv (2\pi)\int^1_0 d\mu \int^{r_{\rm max}}_0 r^2dr \xi(r,\mu, z)  ,
\eea
where $r$ is radial separation of galaxy pair and $\mu$ is the cosine of the angle between the line-of-sight and pair-separation direction.
Then, the normalized 2pCF is
\bea \label{eq:normalize2}
\hat\xi(r,\mu, z) \equiv \frac{\xi(r,\mu, z)}{\xi_*(z)}
\eea
Regarding the choice of $r_{\rm max}=45~h^{-1}~\rm{Mpc}$, we picked a value that is large enough make the normalization insensitive to the highly nonlinear small-scale physics of the Finger-of-God. We find our results are generally insensitive to the choice of $r_{\rm max}$. The left panel of Figure~\ref{fig:xi} shows $\hat\xi$ from one of the sub-cube mock galaxy samples.

In this work, we use the Legendre polynomials $P_0=1$, $P_2=(3\mu^2 - 1)/2$ and $P_4=(35\mu^4 - 30\mu^2 +3)/8$ to approximate the angular-dependence of $\hat{\xi}$ at each $r$:
\bea \label{eq:fit1}
\hat\xi (r,\mu,z) &=& \hat\xi_0 (r,z) P_0 (\mu) \nonumber \\
&&+\hat\xi_2 (r,z) P_2 (\mu) +\hat\xi_4 (r,z) P_4 (\mu).
\eea
Here, $\hat\xi_0$, $\hat\xi_2$, and $\hat\xi_4$ are similar to the monopole, quadrupole, and hexadecapole moments at a given $r$ except that we exclude $r_\perp <3~h^{-1} \rm Mpc$ in the fitting. In this case, we cannot use a decomposition formula like $\hat\xi_\ell = \int \hat\xi(\mu) P_\ell(\mu)d\mu$ and have to make $\chi^2$ fitting instead. We write $\hat\xi_{0,2,4}$ when we refer to $\hat\xi_0$, $\hat\xi_2$, and $\hat\xi_4$ altogether in the rest of the paper. Thanks to this exclusion of the highly non-linear part of the 2pCF, the fit by these 3 moments are highly accurate (Fig.~\ref{fig:xi}). In principle, $\hat{\xi}$ can be decomposed into Legendre polynomials of arbitrary order, but we find that higher order moments do not help much in tightening the constraint. 

The cosmic variance in finite survey volume would give an intrinsic scatter to $\hat\xi_{0,2,4}$. The uncertainty ranges of those moments are computed from the 216 sub-cube mock samples of HR4 and shown as shaded regions in Figure~\ref{fig:xi024}. For our main analysis, we shall use the differences $\hat\xi$ across different redshifts, which measures the shape change of 2pCF between redshift. We shall introduce the difference, $\Delta\hat\xi$, in Sec.~\ref{sec:delxi} along with our motivation for adopting it.

\subsection{Geometrical Distortion Effects due to Choice of Incorrect Cosmology}

If an incorrect cosmology is adopted when converting redshift to distance, then the apparent spatial distribution of galaxies will be distorted. Recalling that the comoving displacements are $\Delta r_\parallel=\Delta z[c/H(z)]$ and $\Delta r_\perp  = (1+z)D_A (z)\Delta \theta$, for an object subtending $\Delta z$ and $\Delta\theta$ in the parallel and perpendicular to the line-of-sight direction, respectively, the distortion in each direction can be parameterized by 
\bea
&&\alpha_\parallel (z) =  \frac{H_{\rm adopted} (z)}{H_{\rm true} (z)}\nonumber \\
&&\alpha_\perp (z) = \left[\frac{D_{\rm A, adopted} (z)}{D_{\rm A, true} (z)}\right]^{-1}.
\eea
In the case of our fiducial cosmology $(\Omega_m,{w})=(0.26,-1)$, if we adopt `incorrectly' that $(\Omega_m,{w})=(0.41, -0.5)$ then $\alpha_\parallel = 0.735$ and $\alpha_\perp = 0.841$ at $z=1$. Relative to the reference point this is a $-26.5$\% and $-15.9$\% change in $r_\parallel$ and $r_\perp$, respectively (see Figure~\ref{fig:slice}), which makes the apparent shape of the galaxy distribution compressed relatively more along the line of sight.

The distortion effect in the 2pCF due to adopting incorrect cosmology has been described by \citet{Li_2016}.
The effect is described by the coordinate transformation \citep{Li_2018}
\bea \label{eq:transformation0}
\xi^\prime(r^\prime_\perp,r^\prime_\parallel) = \xi(r_\perp/\alpha_\perp ,r_\parallel/\alpha_\parallel).
\eea 

In polar coordinates, the transformation is  $\xi^\prime(r^\prime,\mu^\prime) = \xi(r,\mu)$, where
\bea \label{eq:transformation}
&&r= r^\prime \sqrt{\alpha_\parallel^{-2} {\mu^\prime}^2 + \alpha_\perp^{-2} (1-{\mu^\prime}^2)}
 \nonumber\\
&&\mu = \mu^\prime\frac{1}{ \alpha_\parallel \sqrt{\alpha_\parallel^{-2} {\mu^\prime}^2 + \alpha_\perp^{-2}(1-{\mu^\prime}^2)}}.
\eea
We normalize the 2pCF after the transformation as in Equation~(\ref{eq:normalize1}) and (\ref{eq:normalize2}).

Using the transformation, we show how $\hat\xi_{0,2,4}$ is affected by choosing incorrect cosmology. In the left panel of Figure~\ref{fig:xi024}, the black solid lines are $\hat\xi_{0,2,4}$ for the fiducial case of $(\Omega_m,{w}) = (0.26,-1)$ and the other lines are those with the distortion effect applied for four incorrect cosmologies $(\Omega_m,{w}) = (0.26,-0.5)$, $(0.26,-1.5)$, $(0.21,-1)$, and $(0.31,-1)$. The significance of the distortion effect appears strongest for $\hat\xi_2$ at separations roughly between 7 and 15 $h^{-1}$ Mpc. $\hat\xi_2$ in the incorrect cosmologies fall nearly outside the 2-$\sigma$ uncertainty in that range. The distortion effect is smaller for $\hat\xi_0$ and $\hat\xi_4$, but it does seem be significant for certain separations. 

\subsection{Cosmology Dependence in the Shape of 2-point Correlation Function}

If the amount of redshift distortion of $\hat\xi_{0,2,4}$ remained the same for different cosmologies, we would be able to use $\hat\xi_{0,2,4}$ to constrain the cosmology using the corrections for non-linear evolution effects found for just one cosmology. However, $\hat\xi_{0,2,4}$ {\it does} have some cosmology dependence as we shall show in this section.

In the right panel of Figure~\ref{fig:xi024}, we show $\hat\xi_{0,2,4}$ for $(\Omega_m,{w}) = (0.26,-1)$, $(0.26,-0.5)$, $(0.26,-1.5)$, $(0.21,-1)$, and $(0.31,-1)$ at $z=1$, which are calculated from the Multiverse simulation set. In comparison to the left panel where we plotted the effect of incorrect cosmology choice, we are describing the intrinsic cosmology dependence of $\hat\xi_{0,2,4}$ in the right panel. The cosmology dependence is tiny for $\hat\xi_4$, but it is significantly large for $\hat\xi_0$ and $\hat\xi_2$ reaching nearly 2-$\sigma$ level for certain cases.
In the next section it is shown that the redshift evolution of $\hat\xi_{0,2,4}$ also depends on cosmology, and this is what needs to be corrected.

\subsection{Evolution of the Shape of 2-point Correlation Function}\label{sec:delxi}

Our AP method does not care whether or not the amplitude or shape of the correlation function depends on cosmology as it cares only about if the function changes across redshifts or not. 
We shall show that the {\it redshift evolution} of 2pCF, namely $\Delta\hat{\xi}_{0,2,4}(z_i,z_j) \equiv \hat{\xi}_{0,2,4}(z_i) - \hat{\xi}_{0,2,4}(z_j)$ between two different redshifts for example, is not sensitive to the underlying cosmology in this section. 

We compute $\Delta\hat{\xi}_{0,2,4}(z_i,z_j)$ for $z_i=1$ and $z_j=0$ for every possible pair of mocks of the total 216 HR4 mocks excluding the cases that use a same mock for both $\hat{\xi}_{0,2,4}(z_i=1)$ and $\hat{\xi}_{0,2,4}(z_j=0)$. Using the same volume for $z=1$ and 0 would underestimate the scatter in $\Delta\hat{\xi}$ because the cosmic variance will be mostly cancelled out. We thus has $215^3$ cases of $\Delta\hat{\xi}_{0,2,4}(z_i,z_j)$, which we plot in Figure~\ref{fig:delxi024} with the effect of incorrect cosmology choice in the left panel and intrinsic cosmology dependence in the right panel. 

$\Delta\hat{\xi}$ is very small compared to $\hat{\xi}$ (see Fig.~\ref{fig:xi024}). This was also addressed in \citet{Li_2016}. Therefore, even though the shape of the 2pCF itself is significantly distorted due to the redshift-space distortion effect, it is roughly a cosmic invariant and can be used for the AP test. In reality, the shape of the redshift-space 2pCF mildly evolves due to non-linear gravitational evolution and change of type of galaxies. Due to the non-linear effect, $\Delta\hat{\xi}_2$ shows nonzero residual that decrease as $r$.
However, we can see from the left panel of Figure~\ref{fig:delxi024} that this change is smaller compared to that produced by the geometrical distortion caused by adopting an incorrect cosmology \citep[see also][]{Li_2016}. 

The cosmology dependence of $\Delta\hat{\xi}$ appears significantly smaller than that of $\hat{\xi}$.
There is more than $2$-$\sigma$ level of change in $\hat{\xi}_0$ when changing $\Omega_m$ from $0.26$ to $0.21$ or $0.31$ (Fig.~\ref{fig:xi024}), but that in $\Delta\hat{\xi}_0$ is well below the $1$-$\sigma$ uncertainty at most separations. The change in $\hat{\xi}_2$ is at 1 to $2$-$\sigma$ level, but that in $\Delta\hat{\xi}_2$ is mostly below $1$-$\sigma$. Both $\hat{\xi}_4$ and $\Delta\hat{\xi}_4$ do not seem to be affected by background cosmology to significant level.

However, the residual cosmology dependence in $\Delta\hat{\xi}$ is {\it not} negligibly small everywhere and it has to be taken in account. We expect the $1$-$\sigma$ level change in $\Delta\hat{\xi}_2$ at separations $5 \sim 10$ $h^{-1}$ Mpc to significantly affect the results (see middle right panel of Fig.~\ref{fig:delxi024}). That is where we expect to have the strongest constraint from the geometrical distortion effect. We describe how we subtract this effect in Section~\ref{sec:L}.

\subsection{Parametrization of Redshift Evolution of 2pCF} \label{sec:fit}

\begin{figure*}
  \begin{center}
    \includegraphics[scale=0.5]{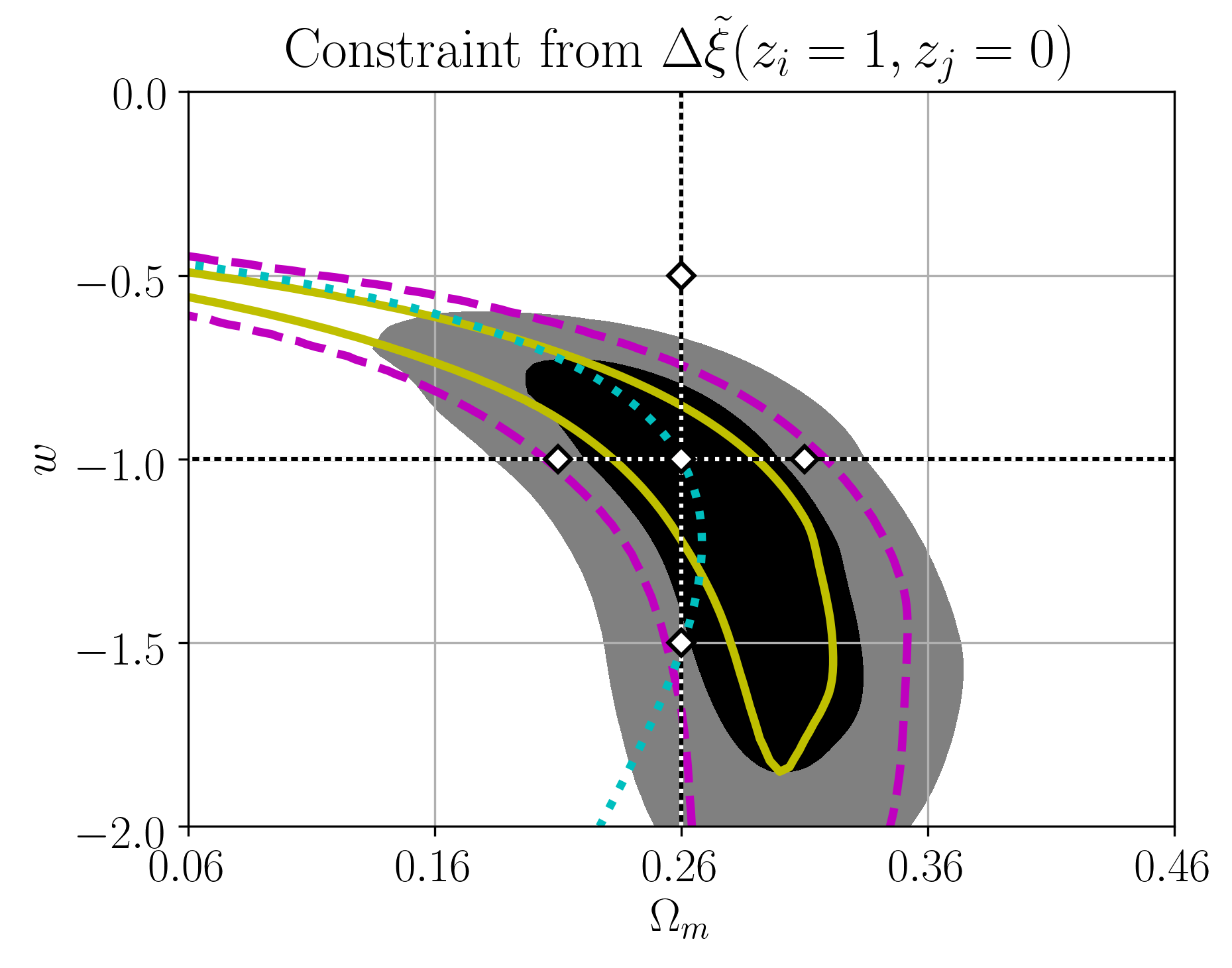}
    \includegraphics[scale=0.5]{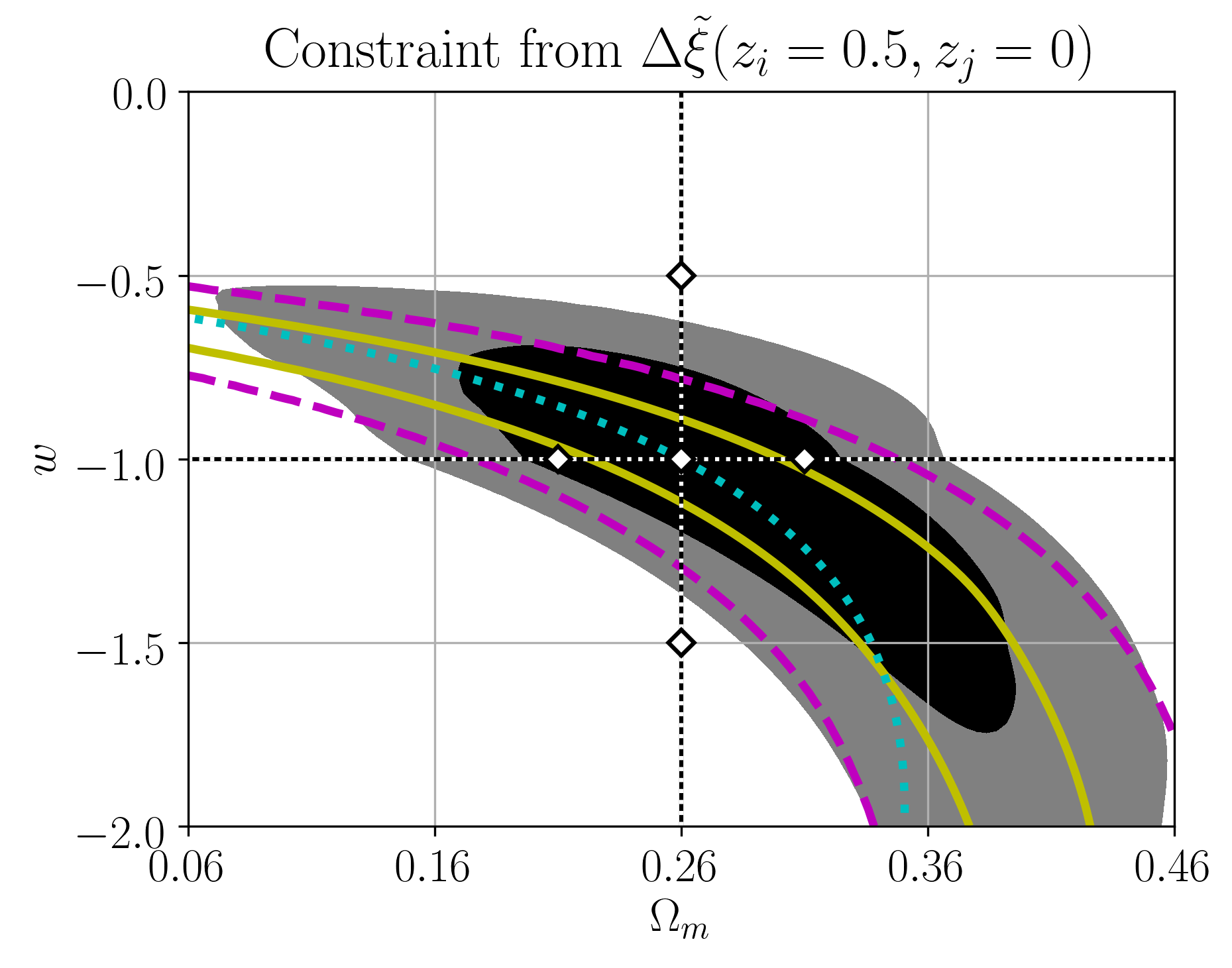}
    \includegraphics[scale=0.5]{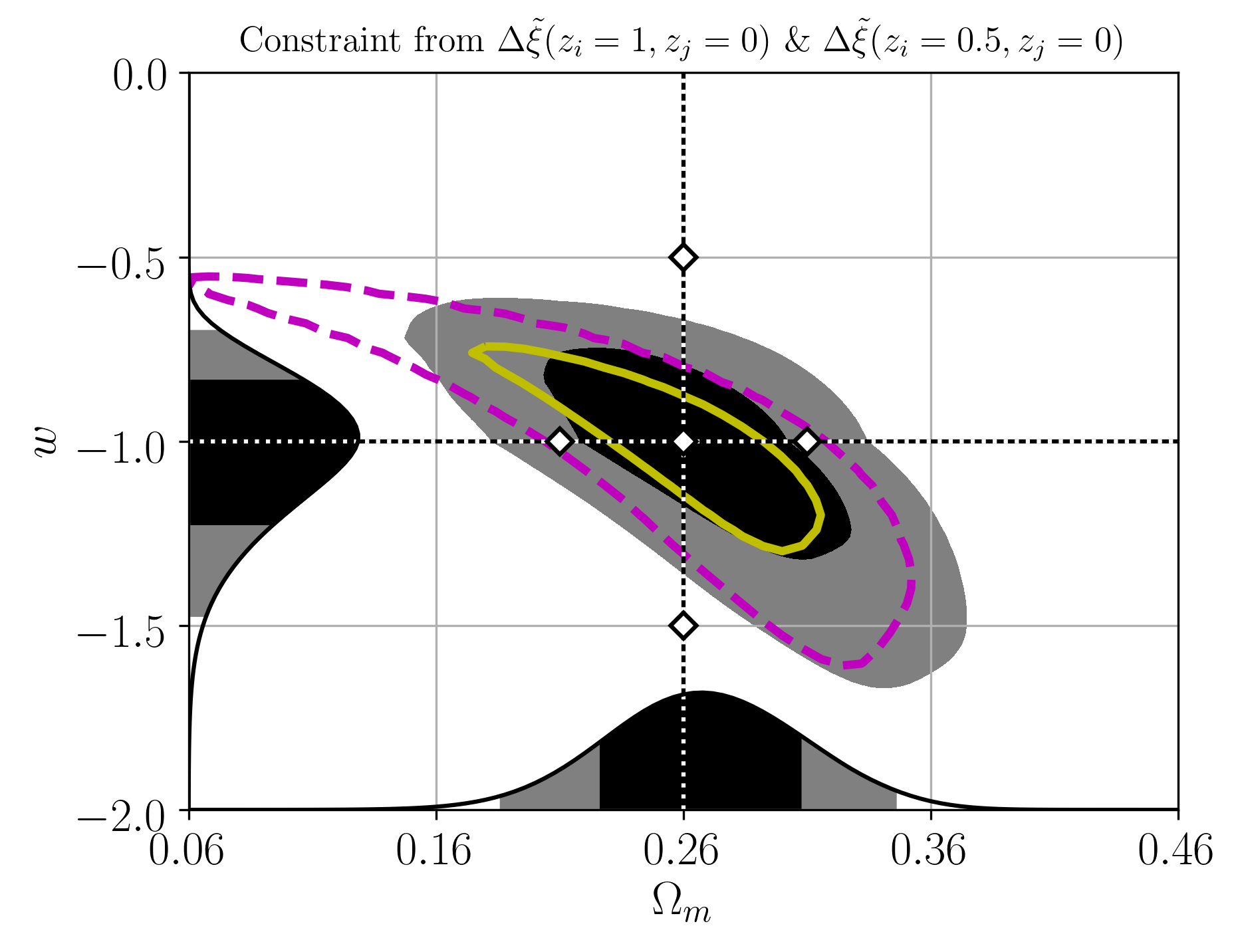}
  \caption{Likelihood map $\mathcal{L}(\Omega_m,{w})$ for the cosmological constraint from $\hat\Delta\xi (z_i=1,z_j=0)$ (upper left), $\hat\Delta\xi (z_i=0.5,z_j=0)$ (upper right) and the combined constraint from both (lower). The black and grey contours enclose the 1 and 2-$\sigma$ ranges of the constraint, respectively. In case of the lower panel, we also plot the marginalized constraint for each parameter as filled curves attached to the axes with the same color convetion for the contours.  The yellow solid and magenta dashed line enclose the 1 and 2$\sigma$ regions, respectively, assuming a fixed systematic correction of $\mathbf{\vec{a}}^{\rm sys}=\mathbf{\vec{a}}_{\rm fid}$ regardless of assumed cosmology. The cyan dotted lines in the upper panels describe the parameter sets that give $\alpha_\perp=\alpha_\parallel$. The diamond symbols denote the five sets of $\Omega_m$ and ${w}$ covered by the multiverse simulations.}
  \label{fig:L}
  \end{center}
\end{figure*}

In order to use $\Delta\hat{\xi}$ for the AP test, we need to compress the information in $\Delta\hat{\xi}$ into a small number of parameters. If the number of the parameters is comparable to that of the samples, the covariance matrix will be significantly biased and the error would propagate to the likelihood evaluation \citep{2007A&A...464..399H, 2014MNRAS.439.2531P}.

We fit the $r$-dependence of the moments, $\Delta\hat\xi_{0,2,4}$, with 2nd order polynomials as follows.
\bea \label{eq:fit2}
&&\Delta\hat\xi_{0,f}(r,z_i,z_j) = r^{-2} \left(a_1 +  a_2 [\log(r)] +  a_3 [\log(r)]^2 \right)\nonumber \\
&&\Delta\hat\xi_{2,f}(r,z_i,z_j) = r^{-2}\left(a_4 +  a_5 [\log(r)] +  a_6 [\log(r)]^2 \right) \nonumber\\
&& \Delta\hat\xi_{4,f}(r,z_i,z_j) = r^{-2} \left( a_7 +  a_8 [\log(r)] +  a_9 [\log(r)]^2  \right)~~
\eea
Above fitting results in 9 parameters $\mathbf{\vec{a}}\equiv(a_1,a_2,...,a_9)$ that describe $\Delta\hat\xi$. Namely,
\bea
&&\Delta\hat\xi \approx \sum_{\ell = 0,2,4} \Delta\hat\xi_{\ell,f}(r) P_\ell(\mu).
\eea
We shall use this 9-element vector $\mathbf{\vec{a}}$ to describe $\Delta\hat\xi$ to calculate likelihood for each cosmology. We find that the constraint is strongest when we fit between $r=5$ and $15~h^{-1}{\rm Mpc}$. $\hat\xi$ at $r>15~h^{-1}$ Mpc does not contribute to the constraint of the cosmology and we exclude it from our analysis. We, thus, use that range to generate $\mathbf{\vec{a}}$.

\subsection{Likelihood Analysis} \label{sec:L}

The redshift difference in the shape of 2pCF, $\Delta\hat\xi$, is much smaller than the shape itself ($\hat\xi$), but it does have a non-zero residual as can be seen from the right panel of Figure~\ref{fig:delxi024}. We shall refer to the value as the {\it systematics} and use the superscript ``$\rm sys$'' to denote it. Our goal is to make an accurate subtraction of $\mathbf{\vec{a}}^{\rm sys}$ to the observed value of $\mathbf{\vec{a}}$ in our likelihood analysis. 

$\Delta\hat\xi_{0,2,4}(z_i,z_j)$ is cosmology dependent as shown in right panel of Fig.~\ref{fig:delxi024}. \citet{Li_2016,Li_2017,Li_2018} assumed it is cosmology independent and it was a potential caveat in their studies. 

In principle, $\bold{a}^{\rm sys}$ should be computed for every cosmology under consideration, which would be too expensive. In this work we linearly interpolate and extrapolate five cases of $\mathbf{\vec{a}}^{\rm sys}$ from the Multiverse simulations with $(\Omega_m,{w})=(0.21,-1)$, $(0.26,-0.5)$, $(0.26,-1)$, $(0.26,-1.5)$, and $(0.31,-1)$. For example, when we compute the likelihood for a given cosmology of $(\Omega_m,{w})=(0.33,-1.1)$, our systematics correction is 
\bea
\mathbf{\vec{a}}^{\rm sys}(\Omega_m,{w}) &=& \mathbf{\vec{a}}^{\rm sys}_{\rm fid} \nonumber \\
&+&(\Omega_m -0.26)
\frac{\mathbf{\vec{a}}^{\rm sys}_{{\rm High-}\Omega_m}- \mathbf{\vec{a}}^{\rm sys}_{\rm fid}}{0.31 - 0.26} \nonumber \\
&+&({w} - (- 1))
\frac{\mathbf{\vec{a}}^{\rm sys}_{{\rm Low-}{w}}- \mathbf{\vec{a}}^{\rm sys}_{\rm fid}}{-1.5 - (- 1)}. \nonumber
\eea
The regions of the parameter space with $\Omega_m<0.21$, $\Omega_m>0.31$, ${w}>-0.5$, or ${w}>-1.5$ requires extrapolation in the systematics, which might be less reliable than interpolation. However, we expect the likelihood for that part of the parameter space to be fairly low and have minimal impact on high-likelihood region. We use the systematics-corrected fitting parameters,
\bea
\mathbf{\vec{p}}(\Omega_m,{w})\equiv \mathbf{\vec{a}}(\Omega_m,{w}) - \mathbf{\vec{a}}^{\rm sys}(\Omega_m,{w})
\eea
for the likelihood calculation.

Then, we calculate the covariance matrix, $\mathcal{C}_{ij}$, using $\bold{p}$ from $216^2$ combinations of HR4 sub-cube mocks at $z=0$ and 1 for the fiducial cosmology $(\Omega_m,{w})=(0.26,-1)$. This matrix describes the uncertainty range of $\Delta\hat\xi$ in the fiducial cosmology. 

Next, we compute $\mathbf{\vec{p}}$ for arbitrary cosmology. For an adopted cosmology, we transform $\xi^{\rm sys}$ using Equation~(\ref{eq:transformation}) and compute $\mathbf{\vec{a}}$ from it. This would put the center of our constraint at the fiducial cosmology $\Omega_m=0.26$ and $w=-1$. Then, $\mathbf{\vec{p}} = \mathbf{\vec{a}} - \mathbf{\vec{a}}^{\rm sys}$ contains the distortion effect from incorrect cosmology choice, but not the cosmic variance. Finally, the likelihood for any adopted cosmology is $\mathcal{L}=\exp{(-\chi^2/2)}$, where
\bea \label{eq:chisq}
\chi^2(\Omega_m,{w}) \equiv \sum_i \sum_j p_i(\Omega_m,{w})\cdot\mathcal{C}_{ij}\cdot p_j(\Omega_m,{w}).
\eea
We shall show the likelihood results from the above equation in the next section.

\section{Results: Cosmological Constraints} \label{sec:results}

\subsection{Constraints from Different Redshift Intervals}

The likelihood contour for $\Omega_m-w$ from our AP test is shown in Figure~\ref{fig:L}. The size of the contour is our prediction for the constraining power of our method. As mentioned above, the center of the constraint is designed to be at the fiducial cosmology. In an actual analysis of observational data from the fiducial cosmology, the center of constraint will be located randomly within the range of uncertainty (i.e. the area enclosed by the contour).

The constraint from the evolution between $z=1$ and 0 ($\Delta\hat{\xi}(z_i=1,z_j=0)$; top left panel) forms a stretched region. The direction of the stretch is similar to the line that satisfy $\alpha_\parallel=\alpha_\perp$ in the parameter space. This is because the AP test loses the constraining power when the ratio of distortions in parallel and perpendicular to the line-of-sight direction, $\alpha_\parallel/ \alpha_\perp$, remains unchanged.

The likelihood contour for $\Delta\hat{\xi}(z_i=0.5,z_j=0)$ (upper right panel of Fig.~\ref{fig:L}) has a similar stretched shape, but is more tilted toward horizontal direction. This is because the distortion factors $\alpha_\parallel$ and $\alpha_\perp$ have different dependence on $\Omega_m$ and ${w}$ at different redshift. At low redshifts $w$ makes an increasingly important role in determining the expansion history, and thus the AP method becomes more sensitive to $w$ with low redshift data.

Due to the difference in the slopes of constraint from $\Delta\hat{\xi}(z_i=1,z_j=0)$ and $\Delta\hat{\xi}(z_i=0.5,z_j=0)$, combining the two data-set tightens the constraint significantly.
We compute the combined constraint using
\bea
\vec{\bold{p}}^{+}\equiv(p^{z_i=0.5}_1,...,p^{z_i=0.5}_{9},p^{z_i=1}_1,...,p^{z_i=1}_{9}), 
\eea
where we simply concatenated $\vec{\bold{p}}$'s from the two redshift pairs, $(z_i,z_j)=(1,0)$ and $(0.5,0)$. We show $\mathcal{L}(\Omega_m,{w})$ from $\vec{\bold{p}}^{+}$ in the lower panel of Figure~\ref{fig:L}. Combining extra redshift data much tightens the constraint, confining the parameters within a 1-$\sigma$ uncertainty of $\Delta \Omega_m \approx 0.04$ or $\Delta {w} \approx 0.2$ when marginalized over $w$ or $\Omega_m$, respectively.

\subsection{Impact of Cosmology Dependence in the Systematics}

In previous works by \citet{Li_2016,Li_2017,Li_2018}, the systematics was modeled from a single background cosmology and the same correction is made for the entire range of cosmology considered. Here, we assess the impact of accounting for the cosmology dependence of the systematics correction, $\mathbf{\vec{a}}^{\rm sys}$. In Figure~\ref{fig:L} we show the cosmological constraints when a fixed systematics correction of $\mathbf{\vec{a}}^{\rm sys}_{\rm fid}$ is used regardless of adopted cosmology (yellow and magenta contours).

The likelihood contour is much more stretched for fixed systematics correction cases, showing stronger degeneracy along the line of $\alpha_\perp=\alpha_\parallel$. For the combined constraint, the shape is more elongated for fixed systematics, but the area of the contour is not much affected: ignoring cosmology dependence of the systematics correction underestimated the uncertainty in the parameter estimation only slightly (20\%). We, however, note that, if the cosmology dependence is not taken into account, the central value of the constraint is likely to shift in the analysis of real observational data by more than 1-$\sigma$ as shown in $\Delta\hat\xi$ across different cosmologies in the right panel of Figure~\ref{fig:delxi024}. Our likelihood results is designed to be centered at the fiducial cosmology.

It may seem the cosmology dependence in the systematics helps to break the degeneracy for parameter sets that are far from our fiducial choice ($\Omega_m=0.26$ and ${w}=-1$). However, the systematics correction outside the coverage of the Multiverse simulations (See diamond symbols in Fig.~\ref{fig:L}) involves less-reliable extrapolation and we cannot make conclusions for those parameters in this work. Ideally, one needs more simulations to extend the coverage. In practice, the result is unlikely to change significantly due to the extra simulations as other cosmological probes like the CMB show that the constrained area will be within the range of Multiverse simulations.

\subsection{Dependence of Number Density and Type of Galaxies}

The constraining power of our AP test is determined by size of the uncertainty in $\Delta\hat{\xi}_{0,2,4}$, which is quantified by the covariance matrix in Equation~(\ref{eq:chisq}). The uncertainty range is also described as the shaded areas in Figure~\ref{fig:delxi024}. The smaller the shades are, the stronger the constraint is. This uncertainty is for our fiducial sample that is (525 Mpc/$h$)$^3$ in volume with $10^{-3}$ galaxies per (Mpc/$h$)$^3$.

The uncertainty is presumably from cosmic variance in finite volume and Poisson noise due to limited number of pairs. In that case, one can reduce the uncertainty by increasing either the survey volume or the number density of sample galaxies. In this section, we show the impact of increasing the sample density by ten times. The denser sample contains about 1.45 million galaxies and includes relatively less massive galaxies.
We note that having such a dense sample from near future surveys is {\it not} practical for redshifts of our interest ($z\gtrsim 0.5$) while we will likely have larger volume than our mock from those surveys. We shall explore the case with enlarged survey volume in future works with larger simulations.

\begin{figure}
  \begin{center}
    \includegraphics[scale=0.55]{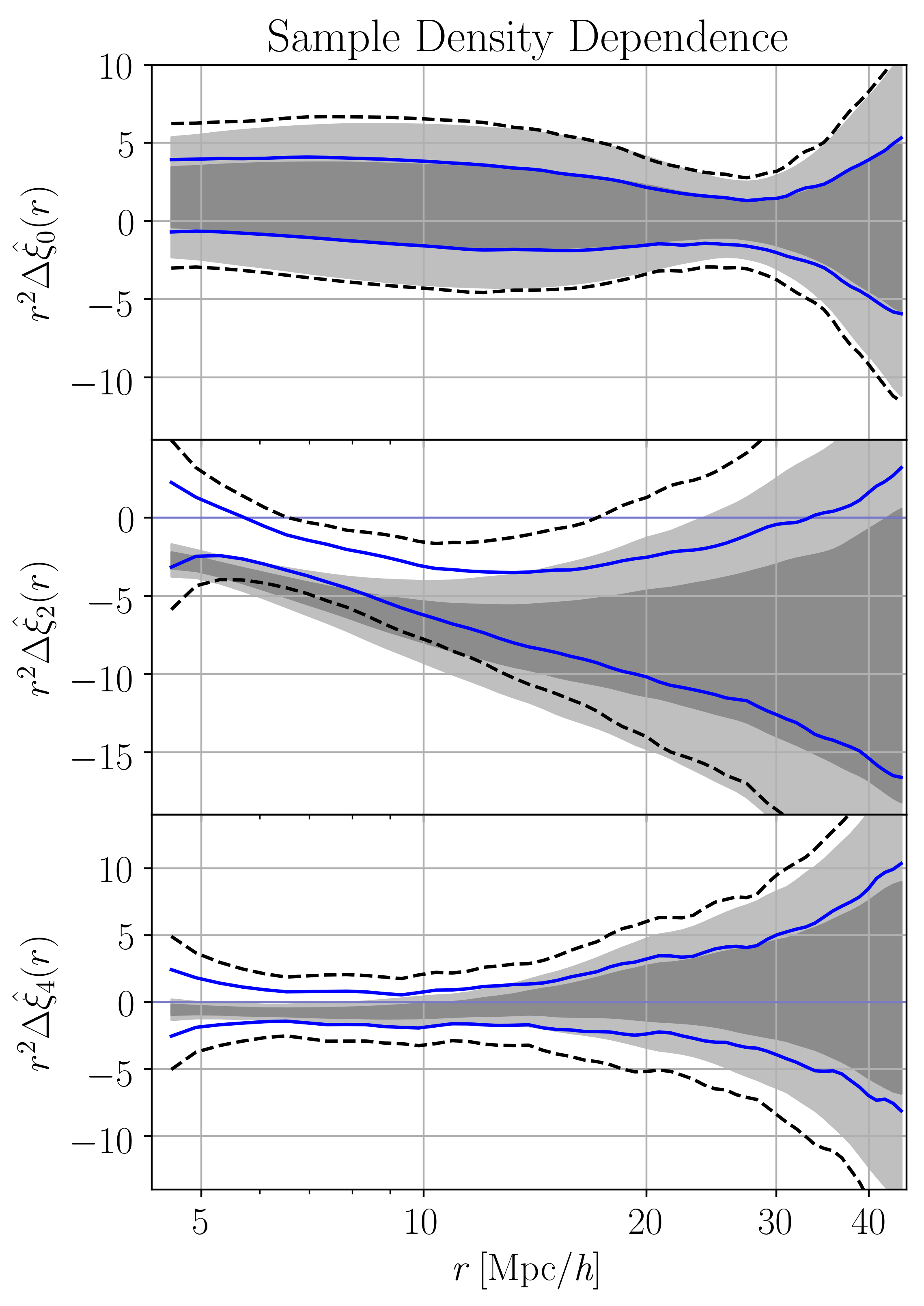}
  \caption{Uncertainty range of $\Delta\hat{\xi}_0(z_i=1,z_j=0)$ (upper panel), $\Delta\hat{\xi}_2(z_i=1,z_j=0)$ (middle panel), and $\Delta\hat{\xi}_4(z_i=1,z_j=0)$ (lower panel) with two different sample galaxy number densities, $n=10^{-2}$ and $10^{-3}~(h^{-1}{\rm Mpc})^{-3}$. The dark and light grey shade describes and 1-$\sigma$ and 2-$\sigma$ uncertainty range, respectively, for mock sample with $n=10^{-2}~(h^{-1}{\rm Mpc})^{-3}$ and the pairs of blue solid and black dashed lines in each panel describes 1-$\sigma$ and 2-$\sigma$ uncertainty range, respectively, for $n=10^{-3}~(h^{-1}{\rm Mpc})^{-3}$.}
  \label{fig:delxi_m2}
  \end{center}
\end{figure}

\begin{figure}
  \begin{center}
    \includegraphics[scale=0.55]{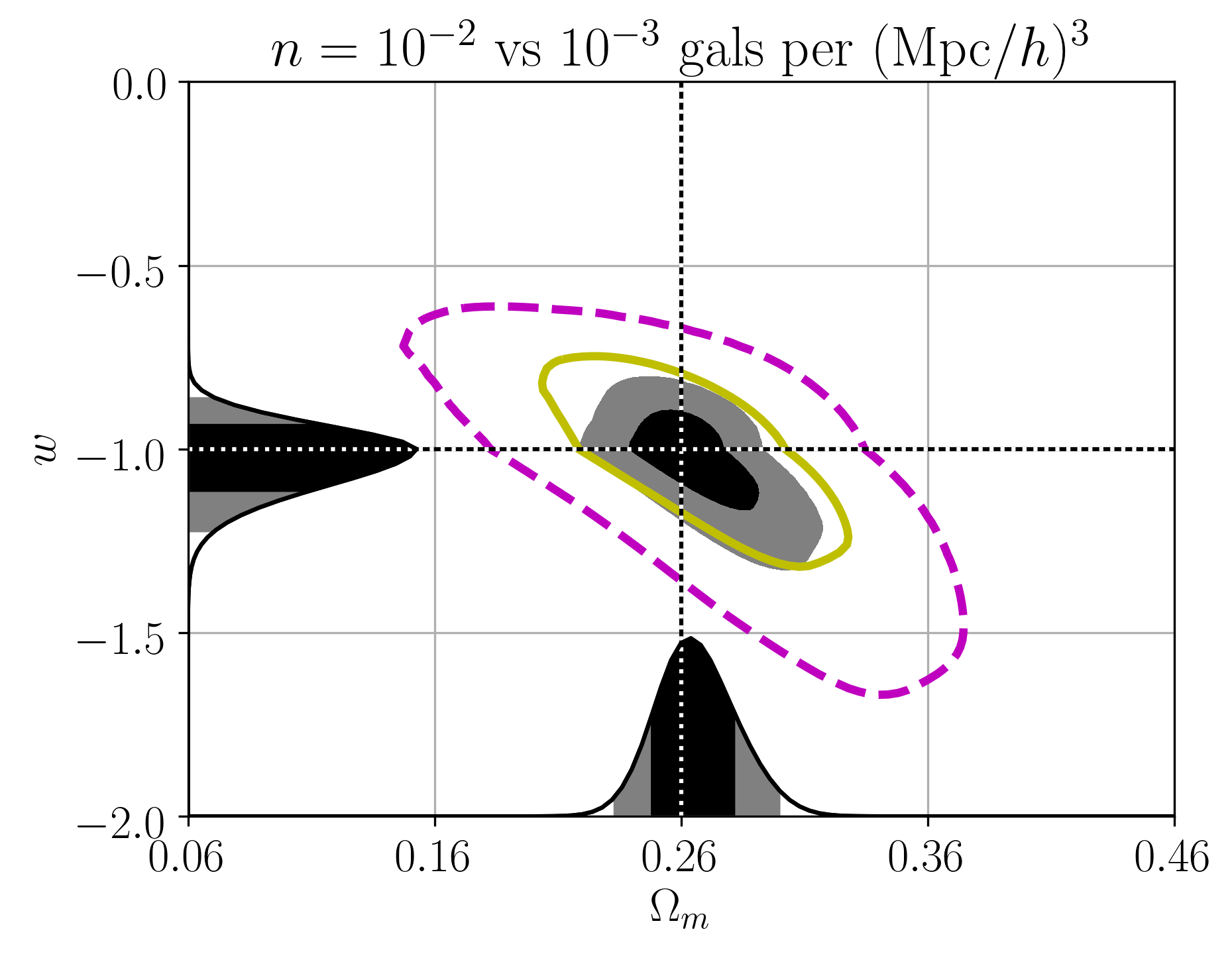}
  \caption{ Constraint from 10 times more galaxies in the same volume. $\mathcal{L}(\Omega_m, w)$ from combined constraint of $\Delta\hat{\xi}(z_i=1,z_j=0)$ and $\Delta\hat{\xi}(z_i=0.5,z_j=0)$ with galaxy number density of $n=10^{-2}~(h^{-1}{\rm Mpc})^{-3}$ is shown in black ($1$-$\sigma$) and grey ($2$-$\sigma$) filled contours. Marginalized constraint for each parameter is shown as filled curves on each axis. The yellow solid and magenta dashed line contours show the constraint from our fiducial galaxy number density of $n=10^{-3}~(h^{-1}{\rm Mpc})^{-3}$, which is same as the black and grey filled contours of Figure~\ref{fig:L}, respectively.}
  \label{fig:L_m2}
  \end{center}
\end{figure}

In Figure~\ref{fig:delxi_m2}, we compare the uncertainty range of our fiducial case $n = 10^{-3}~(h^{-1}{\rm Mpc})^{-3}$ with another sample with ten times higher galaxy number density $n=10^{-2}~(h^{-1}{\rm Mpc})^{-3}$. In the case with higher number density, the uncertainty is substantially reduced for $\Delta\hat{\xi}_2$ and $\Delta\hat{\xi}_4$ at $\lesssim 10~h^{-1}{\rm Mpc}$, where most of the constraint comes from. As a result, the predicted cosmological constraint in the high number density case turns out be substantially tighter. It gives nearly five times smaller 1-$\sigma$ (2-$\sigma$) area and more than two times tighter constraint for each parameter (See Fig.~\ref{fig:L_m2}), giving marginalized constraints of $\Delta \Omega_m \approx 0.017$ or  $\Delta {w} \approx 0.09$ with 1-$\sigma$ uncertainty. Note that this is more than a factor of two improvement compared to the result of the fiducial case, $\Delta \Omega_m \approx 0.04$ \&  $\Delta {w} \approx 0.2$.

Note that average distance between galaxies is 10 ($h^{-1}{\rm Mpc}$) for $n=10^{-3}~(h^{-1}{\rm Mpc})^{-3}$ while it is $\sim4.6$ ($h^{-1}{\rm Mpc}$) for $n=10^{-2}~(h^{-1}{\rm Mpc})^{-3}$. The uncertainty is significantly reduced at $\lesssim 10~h^{-1}{\rm Mpc}$ as we increase the number density while it stays nearly the same at larger scales. The uncertainty seems to be dominated by the cosmic variance on the scales larger than the mean galaxy separation while the shot noise seems to dominate on the scales shorter than the mean separation. Also, it can be seen in Figure~\ref{fig:delxi_m2} that the size of systematics is larger for the less massive galaxies particularly for $\Delta\hat\xi_2$. Therefore, it is necessary to estimate the systematics with the abundance of galaxies matched with observation. 

\subsection{Comparison with the Previous Method} \label{sec:comparison}

\citet{Li_2016,Li_2017,Li_2018} used the binned values of the radially integrated 2pCF, which is similar to
\begin{multline} \label{eq:xiao}
\Delta \hat\xi_{\Delta r}(\mu,z_i,z_j) \equiv \\
\int^{r_{\rm max}}_{r_{\rm min}} \hat\xi(r,\mu,z_i) dr - \int^{r_{\rm max}}_{r_{\rm min}} \hat\xi(r,\mu,z_j) dr,
\end{multline}
where $r_{\rm min}=5~h^{-1}{\rm Mpc}$ and $r_{\rm max}=45~h^{-1}{\rm Mpc}$ are chosen in their AP test. Their method potentially suffers from loss of information in the radial shape of the 2pCF. We note that the radial integrals the right-hand-side of Equation~(\ref{eq:xiao}) is practically dominated by $\xi$ at $r = r_{\rm min}$ because of the $r^{-2}$-like scaling of $\xi$.

We expect the constraining power of the AP test to improve with our method that uses both radial and angular dependence of $\xi$. 
To compare the constraining power of the two methods, we reproduce their AP test with nine $\mu$-bins of $\Delta\hat\xi_{\Delta r}(\mu,z_i=1,z_j=0)$ and with $r_{\rm min}=5~h^{-1}{\rm Mpc}$ and $r_{\rm max}=45~h^{-1}{\rm Mpc}$. We set the number of the $\mu$-bins to be the same as the number of parameters we use to fit $\Delta\hat\xi_{0,2,4}$. The systematics correction is assumed to be cosmology-independent in those works. That is $\mathbf{\vec{a}}^{\rm sys}=\mathbf{\vec{a}}^{\rm sys}_{\rm fid}$ for all cosmologies.

\begin{figure}
  \begin{center}
    \includegraphics[scale=0.55]{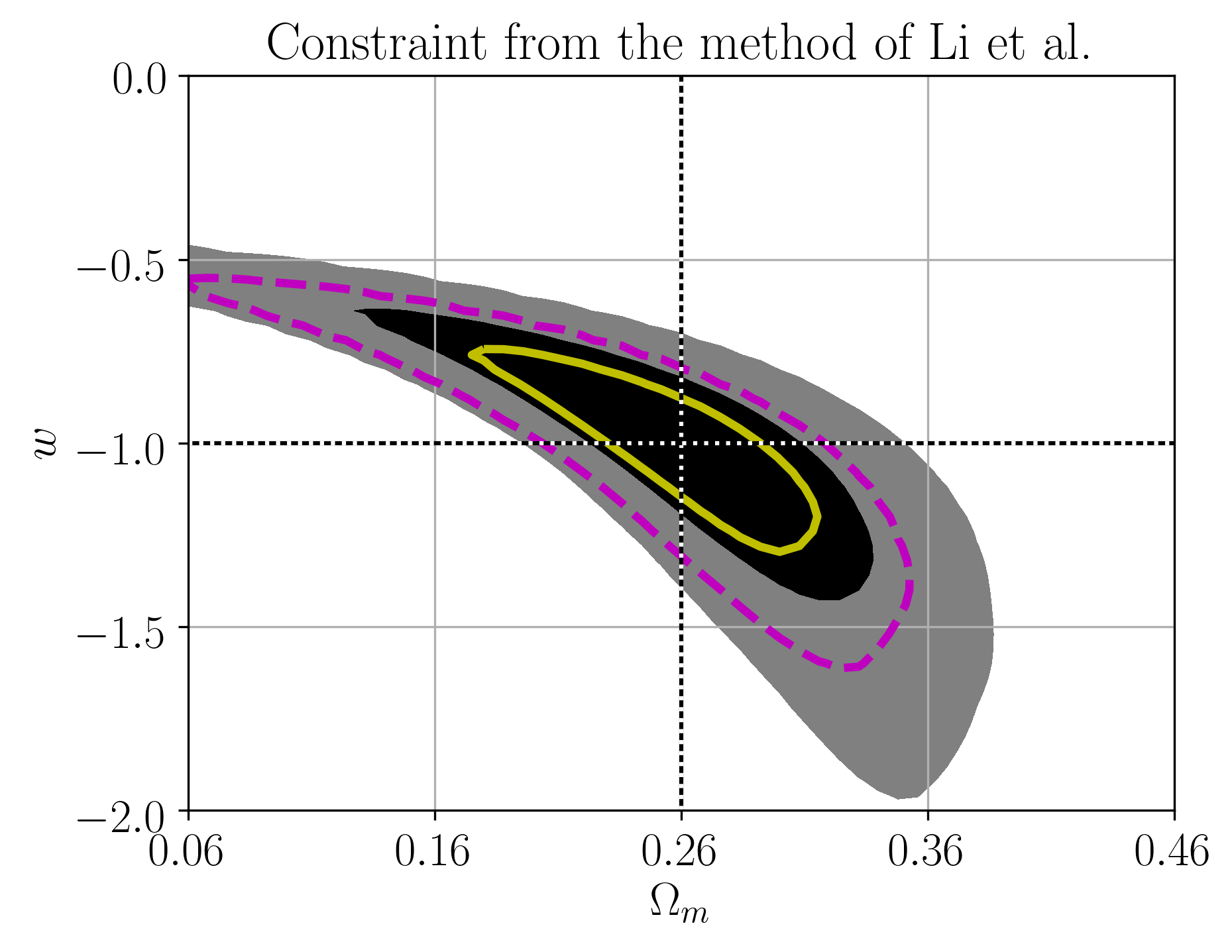}
  \caption{Constraint from the method of \citet{Li_2016}. $\mathcal{L}(\Omega_m, w)$ from combined constraint of $\Delta\hat{\xi}_{\Delta r}(z_i=1,z_j=0)$ and $\Delta\hat{\xi}_{\Delta r}(z_i=0.5,z_j=0)$ from Equation~\ref{eq:xiao} is shown in black ($1$-$\sigma$) and grey ($2$-$\sigma$) filled contours. For comparison, the constraint from this work assuming cosmology independent systematics is shown in yellow solid ($1$-$\sigma$) and magenta dashed ($2$-$\sigma$) contours. Note that the line contours are same as those in the lower panel of Figure~\ref{fig:L}.}
  \label{fig:L_Xiao}
  \end{center}
\end{figure}

The resulting likelihood for $\Omega_m-{w}$ is shown in Figure~\ref{fig:L_Xiao}. The constraint from the method of \citet{Li_2016,Li_2017,Li_2018} gives about 40\% larger uncertainty in the parameter estimation (black and grey areas). Clearly, using the full shape does improve the constraint by a significant amount. 

The amount of improvement in constraint, however, may not look very impressive considering that we added an extra dimension in the analysis. This is because the geometric distortion effect in $\Delta\hat\xi_{0,2,4}$ over different $r$'s is correlated to some extent. As we see in the left panel of Figure~\ref{fig:delxi024}, incorrect cosmology choice shifts or tilts $\Delta\hat\xi_{0,2,4}$ more or less uniformly over $r$. In this case, combining constraints over different $r$'s does not add up perfectly. 
Nevertheless, using the full shape does improve the constraint by a significant amount.

\section{Summary and Conclusions} \label{sec:summary}

The AP test that uses the evolution of redshift-space 2pCF as proposed by \citet{Li_2015} and \citet{Li_2016} is a powerful method for constraining the cosmological parameters governing the expansion of the universe. We presented a new improved method for the AP test that utilizes the two-dimensional shape of the anisotropic galaxy clustering down to a scale as small as $5~h^{-1}{\rm Mpc}$. We also showed the importance of accounting for the cosmology dependence of the systematics correction, which has been neglected in previous works. In this work we focused on describing and justifying the method with ideal mock galaxy samples constructed from a high resolution large volume $N$-body simulation. We shall apply this methodology to observational data in future works. 

Our method decomposes the 2-dimensional galaxy 2pCF into the Legendre polynomials whose amplitudes are modeled by radial fitting functions (Eq.~\ref{eq:fit1}~\&~\ref{eq:fit2}). This allows us to describe the 2-D shape of the 2pCF with a reasonably small number of parameters. Our likelihood analysis with this 2-D fitting scheme tightens the constraint on $\Omega_m$ and ${w}$ by $40\%$ compared to the previous method of \citet{Li_2016,Li_2017,Li_2018} that uses one dimensional angular dependence only. 

We found that the systematic effects in the shape of 2pCF has a non-negligible amount of cosmology dependence over $\Omega_m=0.21$ - $0.31$ and $w=-0.5$ - $-1.5$, which can results in changes in the shape of constraint. The cosmology dependence is likely to change the center of constraint in the case of observational data. Therefore, it would be desirable to account for the cosmology dependence with more simulations from different background cosmologies in future works. 

The constraint on $\Omega_m$ and ${w}$ from a single pair of redshift has a degeneracy for the parameter sets that give the same $\alpha_\perp/\alpha_\parallel$. This degeneracy can be broken by adding extra pair of redshifts in the analysis. 
Most of the constraint comes the smallest scales we consider, which is between $r = 5$ and 10$~h^{-1}{\rm Mpc}$. Reducing the uncertainty in the shape of 2pCF at those scales is the key to tightening the constraint and this can be achieved by increasing the number galaxy pairs or enlarging the survey volume. When we increased the mock galaxy number density from $n=10^{-3}$ to $10^{-2}~(h^{-1}{\rm Mpc})^{-3}$ while the physical size of the sample is fixed to $(525~h^{-1}{\rm Mpc})^3$, the constraint tightened by nearly twice. 
However, enlarging the sample volume is more relevant to the expected outcome of upcoming surveys, which we shall explore in future works.

\section*{Acknowledgement}

The authors thank C. Pichon, H. S. Hwang, E. Komatsu, D. Jeong and T. Sunayama for the helpful comments on this work. 
XDL acknowledges the supported from NSFC grant (No. 11803094).
CGS acknowledges financial support from the National Research Foundation (NRF; \#2017R1D1A1B03034900, \#2017R1A2B2004644 and \#2017R1A4A1015178). SEH was supported by Basic Science Research Program through the National Research Foundation of Korea funded by the Ministry of Education(2018\-R1\-A6\-A1\-A06\-024\-977).

Authors acknowledge the Korea Institute for Advanced Study for providing computing resources (KIAS Center for Advanced Computation Linux Cluster System). This work was supported by the Supercomputing Center/Korea Institute of Science and Technology Information, with supercomputing resources including technical support (KSC-2016-C3-0071) and the simulation data were transferred through a high-speed network provided by KREONET/GLORIAD.

\end{CJK}
\bibliographystyle{apj}
\bibliography{reference}
\end{document}